\documentclass[%
 reprint,
superscriptaddress,
bibnotes,
 amsmath,amssymb,
 aps,
]{revtex4-2}

\usepackage[utf8]{inputenc}
\usepackage{siunitx}
\usepackage{array}
\usepackage{physics}
\usepackage{dsfont}
\usepackage[english]{babel}			
\usepackage{xcolor}
\usepackage{graphicx}				
\usepackage{amssymb,amsmath}		
\usepackage{url}					
\usepackage{marvosym}				
\usepackage[T1]{fontenc}            %
\usepackage{wrapfig}				
\usepackage{charter} 				
\usepackage{blindtext}				
\usepackage{datetime}				
\usepackage{lipsum}                 
\usepackage{float}                  
\usepackage{tabularx}
\usepackage{dcolumn}
\usepackage{bm}
\usepackage{todonotes}
\usepackage{hyperref}
\hypersetup{colorlinks=true,allcolors=blue}

\newcommand{\UIBK}{Institut f{\"u}r Experimentalphysik, Universit{\"a}t Innsbruck, 6020 Innsbruck, Austria}
\newcommand{\WWU}{Institut f{\"u}r Festk{\"o}rpertheorie, Universit{\"a}t M{\"u}nster, 48149 M{\"u}nster, Germany}
\newcommand{\TUdo}{Condensed Matter Theory, Department of Physics, TU Dortmund, 44221 Dortmund, Germany}
\newcommand{\JKU}{Institute of Semiconductor and Solid State Physics, Johannes Kepler University Linz, 4040 Linz, Austria}
\newcommand{\ACP}{Institute of Applied Physics, Abbe Center of Photonics, Friedrich Schiller University Jena, 07745 Jena, Germany}
\newcommand{\IOF}{Fraunhofer Institute for Applied Optics and Precision Engineering IOF, Center of Excellence in Photonics, 07745 Jena, Germany}

\preprint{APS/123-QED}

\begin{document}

\title{Compact Chirped Fiber Bragg Gratings for Single-Photon Generation from Quantum Dots}

\author{Vikas Remesh}
\thanks{These authors contributed equally}
\affiliation{\UIBK}
\author{Ria G. Krämer}
\thanks{These authors contributed equally}
\affiliation{\ACP}
\author{René Schwarz}
\thanks{These authors contributed equally}
\affiliation{\UIBK}
\author{Florian Kappe}%
\affiliation{\UIBK}
\author{Yusuf Karli}%
\affiliation{\UIBK}
\author{Malte Per Siems}
\affiliation{\ACP}
\author{Thomas K. Bracht}%
\affiliation{\WWU}
\affiliation{\TUdo}
\author{Saimon Filipe Covre da Silva}%
\affiliation{\JKU}
\author{Armando Rastelli}%
\affiliation{\JKU}
\author{Doris E. Reiter}%
\affiliation{\TUdo}
\author{Daniel Richter}%
\affiliation{\ACP}
\author{Stefan Nolte}%
\affiliation{\ACP}
\affiliation{\IOF}
\author{Gregor Weihs}%
\affiliation{\UIBK}

\date{Date: \today \\ \phantom{XXX} E-mail: vikas.remesh@uibk.ac.at}
\begin{abstract}
A scalable source of single photons is a key constituent of an efficient quantum photonic architecture. To realize this, it is beneficial to have an ensemble of quantum emitters that can be collectively excited with high efficiency. Semiconductor quantum dots hold great potential in this context, due to their excellent photophysical properties. Spectral variability of quantum dots is commonly regarded as a drawback introduced by the fabrication method. However, this is beneficial to realize a frequency-multiplexed single-photon platform. Chirped pulse excitation, relying on the so-called adiabatic rapid passage, is the most efficient scheme to excite a quantum dot ensemble due to its immunity to individual quantum dot parameters. Yet, the existing methods of generating chirped laser pulses to excite a quantum emitter are bulky, lossy, and mechanically unstable, which severely hampers the prospects of a quantum dot photon source. Here, we present a compact, robust, and high-efficiency alternative for chirped pulse excitation of solid-state quantum emitters. Our simple plug-and-play module consists of chirped fiber Bragg gratings (CFBGs), fabricated via femtosecond inscription, to provide high values of dispersion in the near-infrared spectral range, where the quantum dots emit. We characterize and benchmark the performance of our method via chirped excitation of a GaAs quantum dot, establishing high-fidelity single-photon generation. Our highly versatile chirping module coupled to a photon source is a significant milestone toward realizing practical quantum photonic devices.  

\end{abstract}

\maketitle

\section*{Introduction}
Future quantum communication architectures rely on bright and efficient sources of high-purity single-photons and entangled photon pairs with superior photostability and scalability. Semiconductor quantum dots are the most important candidates for this role, due to their high brightness \cite{senellart_high-performance_2017,tomm_bright_2021}, high degrees of entanglement \cite{jayakumar_time-bin_2014,schimpf_quantum_2021}, low multiphoton rate \cite{hanschke_quantum_2018} and near-deterministic operating nature. 
An important feature of quantum dots is the possibility to control the emission wavelength of an ensemble by properly choosing  materials, growth methods, and fabrication parameters. Further post-growth tuning is offered by electric \cite{gerardot2007manipulating}, magnetic\cite{koong_multiplexed_2020}, optical \cite{grim_scalable_2019} or strain field tuning \cite{trotta2012universal}. However, these methods increase the complexity of the photon source. The natural spread in the dots' physical properties like size, composition, and strain is commonly perceived as an obstacle to achieving scalable quantum hardware. However, the resulting distribution in emission wavelengths offers a convenient way to produce frequency-multiplexed single-photons and entangled photon pairs. Therefore, it is important to develop a simple, efficient, and long-term stable optical excitation method that delivers high-quality photons from the source. 

Several resonant and off-resonant optical excitation schemes have been developed to generate single-photons or entangled photon pairs from quantum dots \cite{matthiesen2012subnatural,benson2000regulated, reindl_phonon-assisted_2017,koong2021coherent,karli2022super,wilbur2022notch,sbresny2022stimulated,wei2022tailoring}. In particular, chirped excitation relying on Adiabatic Rapid Passage (ARP) is a robust scheme for quantum dot excitation as demonstrated before \cite{wu_population_2011,glassl_biexciton_2013,debnath_high-fidelity_2013,mathew2014subpicosecond,kaldewey_coherent_2017,kappe2022collective}. 
The ARP excitation is immune to fluctuations in the laser power and spectral properties of quantum dots, which facilitates a collective driving to produce single-photons \cite{ramachandran_experimental_2021,kappe2022collective} or entangled photon pairs \cite{creatore_creation_2012}. 
It is also known that in ARP, the excitation efficiency depends on the sign of the chirp, i.e., positive chirp achieves near-unity excitation efficiency, while for negative chirp, preparation efficiencies are lower due to phonon influences \cite{luker_influence_2012}. 
\begin{figure*}[htb!]
	\centering
	\includegraphics[width=1\linewidth]{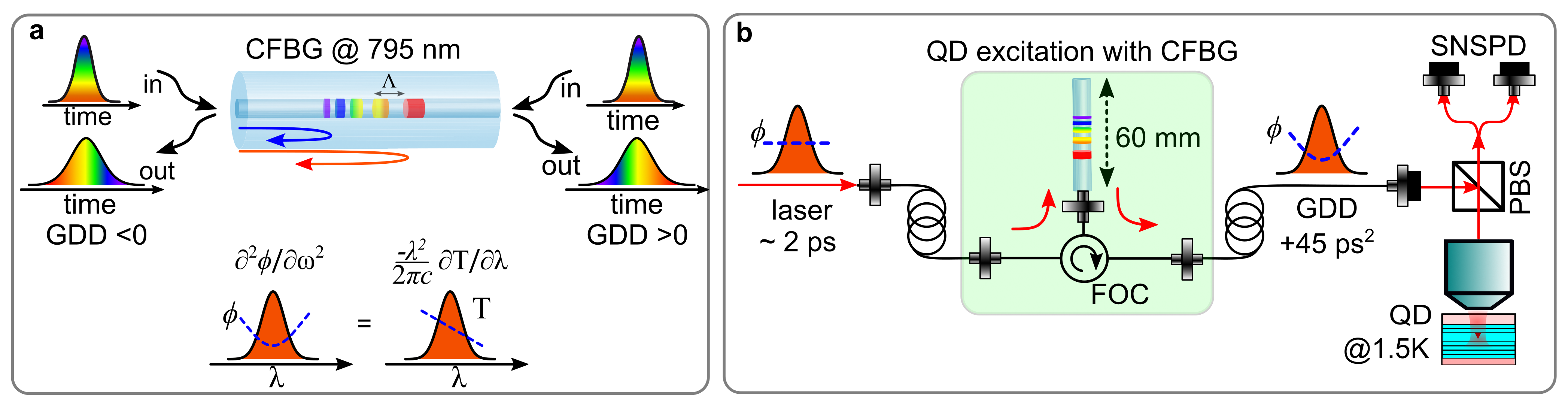}
	\caption{\textbf{Plug-and-play chirping module for quantum dots}: (a) Working principle of a CFBG for pulse chirping. The direction of laser pulse propagation through the CFBG dictates the sign of the group delay dispersion (GDD): entry and reflection from the left side results in negatively chirped pulses and vice versa. The relation between the GDD and wavelength-dependent time delay ($T$) is also illustrated. (b) Single-photon generation: \SI{2}{\pico\second}-long laser pulses, centred at \SI{794.5}{\nano\meter} are coupled to the CFBG to impart \SI{45}{\pico\second\squared} chirp. A fiber-optic circulator (FOC) helps separate incident (unchirped) and reflecting (chirped) laser pulses. The chirped laser pulses are directed to an optical cryostat where the quantum dot is mounted. A polarizing beamsplitter (PBS) enables cross-polarization rejection of scattered laser light. The generated single photons at exciton (X) and biexciton (XX) energies are collected via the same path, and filtered and detected using superconducting nanowire single photon detectors (SNSPD).}  
 \label{fig_intro}
\end{figure*} 

In most of the practical implementations, chirped pulses are prepared by introducing the Group Delay Dispersion (GDD) in the laser spectral phase \cite{monmayrant2010newcomer}. Examples include diffraction-grating stretchers \cite{treacy1969optical,martinez_3000_1987,lai_single-grating_1994,pessot19871000,kaldewey_coherent_2017,wei_deterministic_2014,kappe2022collective} and programmable spatial light modulators (SLM) 
\cite{assion1996coherent,weiner2000femtosecond,remesh2019coherent,mathew2014subpicosecond,wilbur2022notch}.
While the SLM-based techniques are user-friendly and enable arbitrary spectral phase manipulation, they are pixellated devices with limited resolution and response times and bear phase wrapping issues. Diffraction-grating stretchers, on the other hand, can be designed to provide extremely high dispersion \cite{banks2000novel}. Yet both of these approaches are bulky, beam-distorting, lossy, and inherit spatiotemporal coupling artifacts and higher-order dispersions \cite{osvay2004angular,sussman2008focusing}. 
To compensate for optical aberrations and higher-order dispersions, one requires rigorous alignment procedures which restrict practical applications. For instance, to accomplish a robust, and optomechanically stable single-mode laser coupling to a quantum dot source, we must ensure a fast and effective approach without spatiotemporal artifacts and aberrations.  

In this Letter, we demonstrate a simple, compact, plug-and-play, and alignment-free approach for chirped pulse excitation of quantum emitters, which circumvents the obstacles in other existing methods. Our design is based on a chirped fiber Bragg gratings (CFBG), fabricated via a femtosecond phase mask inscription, to provide tailored values of GDD around \SI{800}{nm} with large spectral bandwidth and high efficiency. Our results pave the way for realizing compact solid-state single-photon sources in an advanced quantum photonic architecture.

\section*{Chirped fiber Bragg gratings for quantum dot excitation}
A fiber Bragg grating consists of an optical fiber with a periodic variation of the refractive index in the core, providing a narrowband in-fiber reflection. The reflecting wavelength $\lambda_\text{B}$ depends on the grating period $\Lambda$ and the effective refractive index $n_{\text{eff}}$ of the guided mode in the fiber core. In a CFBG, however, the grating period is not constant but is varied along the propagation axis $z$. This results in the reflection of different wavelengths at different spatial positions of propagation:
\begin{equation}
    \lambda_\text{B}(z)=2 n_{\text{eff}}\Lambda(z).
\end{equation}
Hence, different parts of the laser spectrum that is reflected at the CFBG acquire different time delays, i.e., the laser pulse gets chirped in time. In Figure \ref{fig_intro}\textbf{a} we summarize the working principle of a CFBG (see also SI). Note that by adjusting the grating period along the propagation axis, the dispersive response of the CFBG is defined: a linear variation of the grating period introduces second-order dispersion, a quadratic variation introduces third-order dispersion, and so on \cite{imogore2020dispersion}.
Initially proposed for dispersion cancellation in optical telecommunication waveguides \cite{ouellette1987dispersion,hill1994chirped}, CFBGs have been employed for chirped pulse amplification \cite{imeshev2004chirped,caucheteur2010experimental}, pulse shaping \cite{eggleton1994experimental,wang2009fourier,li2011all,imogore2020dispersion} and high time resolution spectroscopy \cite{davis2017pulsed,sosnicki2023interface}, among many others. Yet despite the progress in this technology, CFBGs have not been explored for chirped excitation of quantum emitters around \SI{800}{\nano\meter}.  

\begin{figure*}[hbt!]
	\centering
	\includegraphics[width=1\linewidth]{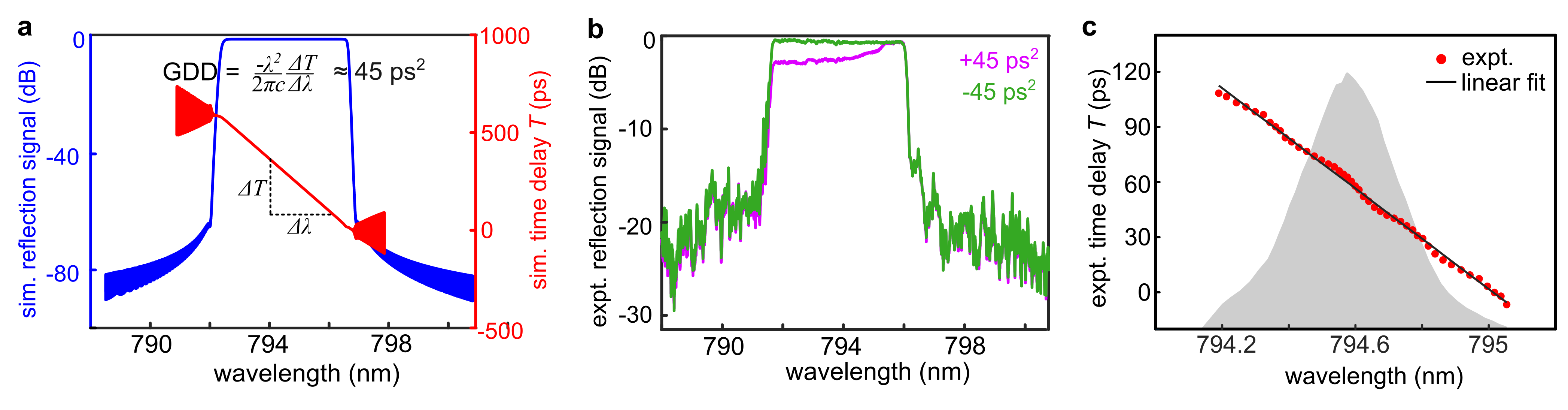}
	\caption{\textbf{Performance of the CFBG}: (a) Simulated reflection spectrum (blue curve) of the CFBG designed for $\pm \SI{45}{\pico\second\squared}$ chirp, centered at \SI{794.5}{\nano\meter} and simulated time delay ($T$, red curve) of different wavelengths. The negative slope of the red curve determines the dispersion parameter $D_{1\lambda}$, from which the GDD is computed. (b) Measured reflection spectra of CFBGs along the +\SI{45}{\pico\second\squared} (magenta curve) and \SI{-45}{\pico\second\squared} (green curve) propagation directions (c) CFBG dispersion measurement: measured time delays ($T$) of different wavelengths via a time-of-arrival technique. Red dots denote the peak wavelengths recorded in the +\SI{45}{\pico\second\squared} direction and the black line denotes the fit. The slope of the solid black line provides the dispersion parameter $D_{1\lambda}$. Grey shade represents the excitation laser spectrum.} 
  \label{fig_cfbgresult}
\end{figure*}

For single-photon generation from quantum dots, CFBGs have a huge potential, as illustrated in Figure \ref{fig_intro}\textbf{b}. To start with, CFBGs are an important breakthrough towards realizing a direct fiber-coupled, alignment-free, high-efficiency, and scalable quantum dot photon source \cite{northeast2021optical,cadeddu2016fiber,zolnacz2019method,gao2022quantum,daveau2017efficient}. They can be tailored to provide arbitrary values of GDD, with a spectral bandwidth of \SI{10}{\nano\meter} or more, coincident with the typical spectral variability of a quantum dot ensemble \cite{da2021gaas,laferriere2022unity}. More importantly, CFBG-based dispersion is robust against mechanical instabilities and vibrations. As it does not rely on angular dispersion, CFBGs eliminate optical aberrations and spatiotemporal couplings, to provide excellent single-mode coupling to the quantum dot source. Furthermore, a diffraction-grating stretcher \cite{martinez_3000_1987,lai_single-grating_1994,backus_high_1998} designed for \SI{45}{\pico\second\squared} chirp is bulky (several meters long, see SI for details). In comparison, a CFBG for the same GDD is only a few millimeters long. The versatility of a CFBG is that it allows inverting the GDD by switching the propagation direction, while in a grating stretcher, adding the option to invert the GDD increases the footprint of the setup even further. Interestingly, with CFBGs we can realize additive combinations \cite{wang2009fourier}, i.e., one can cascade several CFBGs to obtain a range of GDDs, without increasing the complexity of the setup. 

\section*{Experimental implementation}
To inscribe the CFBGs we employ a phase mask scanning technique using femtosecond laser pulses \cite{thomas2012femtosecond} with \SI{800}{\nano\meter} wavelength, \SI{100}{\femto\second} pulse duration at \SI{200}{\hertz} repetition rate (Spectra-Physics Spitfire Ace). The fiber core refractive index modification is induced by multiphoton absorption of the incident laser irradiation. The non-periodic modulation of the refractive index modification is generated by a phase mask (i.e., a transmission grating, manufactured in-house at the University of Jena), which defines the resulting distribution of the grating period along the fiber axis. 
To spectrally characterize the fabricated CFBGs in reflection and transmission, we use linearly polarized light from a supercontinuum source (NKT, SuperK EXR-15) and an optical spectrum analyzer (Yokogawa, OSA AQ6375). A fiber-optic circulator (PMOptics) helps extract the reflected signal of the CFBG. From the transmitted signal, we determine the reflection strength as well as any broadband losses introduced by the inscription process. For dispersion characterization of the CFBG, we rely on a time-of-arrival technique. 

For quantum dot excitation, we tune the laser source to \SI{794.5}{nm}, such that a two-photon resonant excitation to the biexciton state is enabled \cite{jayakumar_time-bin_2014,reindl_phonon-assisted_2017,kaldewey_coherent_2017,kappe2022collective}. The laser beam is directly coupled to a CFBG and the resulting, dispersed pulses are directed to the quantum dot sample (see SI and Ref. \cite{kappe2022collective} for details) mounted in a closed-cycle cryostat with a base temperature of \SI{1.5}{\kelvin} (ICEOxford). 
The quantum dot emission is collected via the same path, and exciton (X) and biexciton (XX) emission photons are filtered using notch filters (BNF-805-OD3, FWHM \SI{0.3}{\nano\meter}, Optigrate) and directed to superconducting nanowire single-photon detectors (SNSPD, Eos, Single Quantum). In the setup we employ cross-polarization filtering for efficient laser scattering rejection.

\section*{Results}
\subsection{CFBG results}
For the experiments described in this paper, we realize a set of three CFBGs with different GDDs (\SI{10}{\pico\second\squared}, \SI{22.5}{\pico\second\squared}, and \SI{45}{\pico\second\squared} respectively). The phase masks employed have their center period around \SI{1090}{\nano\meter}, resulting in CFBG center period of \SI{545}{\nano\meter}, which corresponds to a second-order reflection at \SI{794.5}{\nano\meter} (see SI for details). To calculate the required linear period variation, i.e., the chirp rate of the CFBG, we use the transfer matrix algorithm as described in Ref. \cite{erdogan1997fiber}. The achievable bandwidth of the CFBG is limited by its maximum length, which is \SI{60}{\milli\meter} in our setup. In Figure \ref{fig_cfbgresult}\textbf{a} we present the simulated reflection spectrum (blue curve) and the wavelength-dependent time delay ($T$, red curve) of the \SI{45}{\pico\second\squared} CFBG. From the slope of the red curve, we compute the dispersion parameter $D_{1\lambda}$, from which we obtain the GDD (see SI for details).  

In Figure \ref{fig_cfbgresult}\textbf{b} we present the measured reflection spectra for the $\pm\SI{45}{\pico\second\squared}$ CFBG (magenta and green curves respectively). We observe a reflectivity  $>\SI{90}{\percent}$ from \SI{792}{\nano\meter} to \SI{796}{\nano\meter}, showing an excellent agreement with the simulation in Figure \ref{fig_cfbgresult}\textbf{a}. The average sideband suppression is $\approx\SI{25}{dB}$. Note that, for the $+\SI{45}{\pico\second\squared}$ direction, lower wavelengths experience more losses due to the cladding mode coupling \cite{thomas2012femtosecond} which arises from the chosen inscription pattern.
\begin{figure*}[hbt!]
	\centering
	\includegraphics[width=1\linewidth]{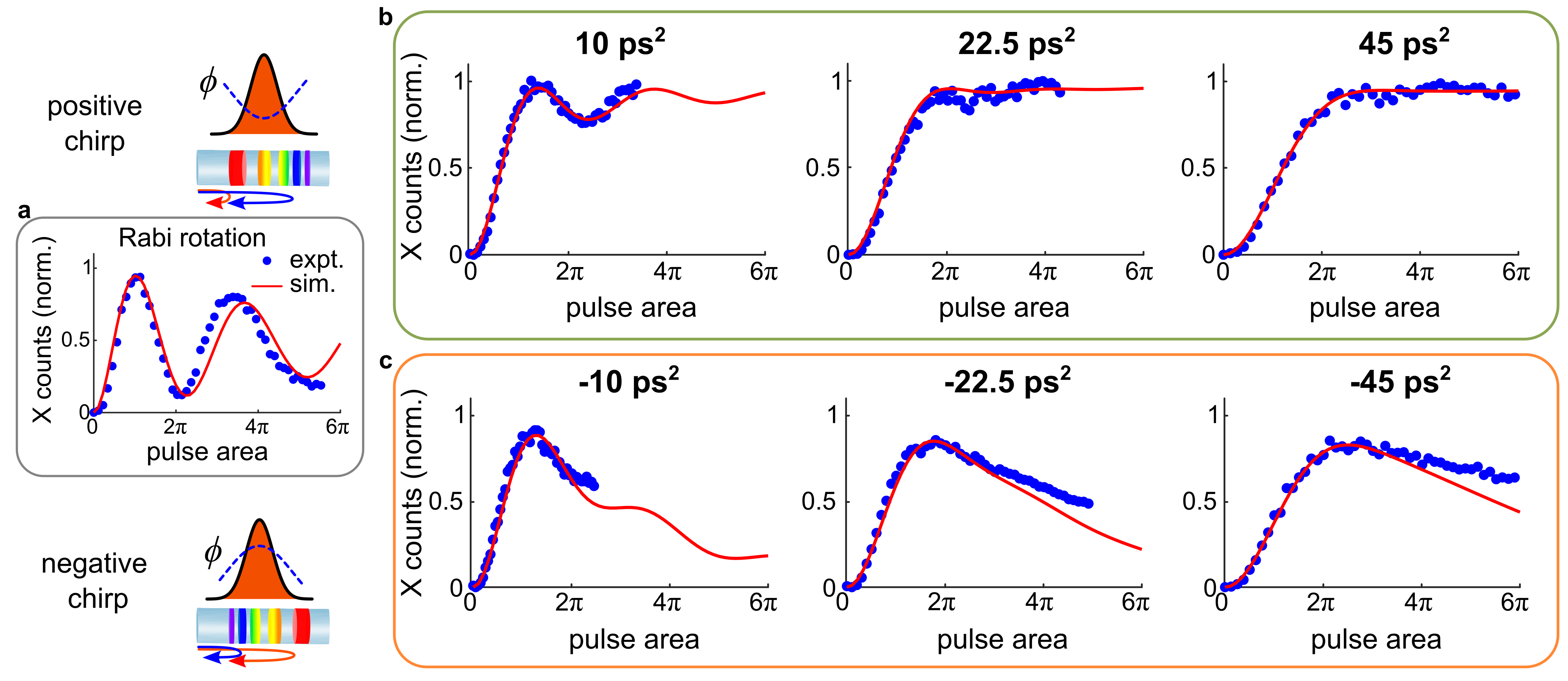}
	\caption{\textbf{Chirped excitation of a quantum dot using CFBG}: (a) the recorded exciton (X) photon counts (blue dots) as a function of pulse area, for the resonant Rabi rotation (i.e., \SI{0}{\pico\second\squared}), (b) positively-chirped excitation, and (c) negatively-chirped excitation. Chirp values are displayed along with. Red curves denote the simulated photon counts for the respective cases.} 
  \label{fig_arp}
\end{figure*}
We attribute a nominal 10\% loss to the absorption and scattering of the guided light in the grating structure. Two other major factors for the loss are the circulator and the connector losses, due to the mismatch of different core diameters (for more details see SI). The latter can be avoided by directly splicing the CFBG to the circulator. Losses due to the splicing of the bare fiber with the inscribed CFBG and connectorized fibers are negligible. 

To characterize the GDD response of CFBGs we perform a time-of-arrival measurement which directly measures the wavelength-dependent time delay (see SI for details). In Figure \ref{fig_cfbgresult}\textbf{c} we display the result for +\SI{45}{\pico\second\squared} CFBG. Based on the variation of the fitted peak wavelengths with respect to their arrival times recorded with the SNSPD, we obtain a $D_{1\lambda} = \SI{-136.77}{\pico\second\per\nano\meter}$, which corresponds to $\text{GDD} =\SI{45.9}{\pico\second\squared}$, showing an excellent match with the simulation in Figure \ref{fig_cfbgresult}\textbf{a}.

\subsection{QD experiment results} 
We then proceed to investigate the single-photon generation from quantum dots. We select a quantum dot resonant within the CFBG bandwidth and characterize it under two-photon excitation (see SI for details). The exciton (X) wavelength is found to be \SI{793.6}{\nano\meter} and the biexciton (XX) wavelength is \SI{795.4}{\nano\meter}. To confirm the resonance condition, we perform a Rabi rotation experiment upon the sweep of the laser power, while recording the corresponding emission spectra. The integrated intensities corresponding to the X emission (for simplicity, XX photon counts are not displayed) are displayed in Figure \ref{fig_arp}\textbf{a}, clearly demonstrating the Rabi oscillations. 

We then investigate the utility of CFBG for chirped excitation of the quantum dot i.e., ARP experiment. To this end, we perform the laser power sweep experiments on the same quantum dot using the three CFBGs (\SI{10}{\pico\second\squared}, \SI{22.5}{\pico\second\squared} and \SI{45}{\pico\second\squared} respectively). The results are displayed in Figure \ref{fig_arp}\textbf{b}, with respective chirp values labeled alongside. At \SI{10}{\pico\second\squared}, we observe that the oscillatory behavior in the recorded X photon counts diminishes. The same experiment at \SI{22.5}{\pico\second\squared} shows that the Rabi oscillations have nearly vanished, and a plateau is attained for higher pulse power, signifying that the process is closer to ARP condition. Finally, with the \SI{45}{\pico\second\squared} CFBG, we observe a clear plateau in the X photon counts for pulse powers >2$\pi$, demonstrating that the ARP process is induced, confirming the previous observations \cite{glassl_biexciton_2013,mathew2014subpicosecond,kaldewey_coherent_2017,wei_deterministic_2014,ramachandran_experimental_2021,kappe2022collective}. 

To further validate the robustness of ARP excitation, we investigate the excited state preparation efficiency under both conditions via a biexciton-to-exciton cross-correlation experiment, as in Ref. \cite{wang2019demand}. We observe that the preparation efficiencies are 93.5\% in the case of Rabi rotation (largest value, i.e., at $\pi$ power) and >94.4\% for ARP condition (for all powers >$\SI{2.5}{\pi}$), in agreement with the previous reports \cite{kappe2022collective,schimpf_quantum_2021} on a similar quantum dot system (for details see SI). Note that, the preparation efficiency under Rabi rotation is extremely sensitive to the pulse power \cite{kappe2022collective}, while under chirped pulse excitation, the preparation efficiency is insensitive to power variations. 
 
To demonstrate the versatility of CFBGs for chirped pulse excitation, we switch the pulse propagation direction (i.e., to obtain a negatively chirped pulse) and repeat the experiments, and the results are displayed in Figure \ref{fig_arp}\textbf{c}. We observe that, as expected \cite{glassl_biexciton_2013,luker_phonon_2017}, the quantum dot excitation efficiency does not reach unity, and the X photon counts drop to 30\% at higher pulse areas, in agreement with earlier experimental observations \cite{mathew2014subpicosecond,kaldewey_coherent_2017}. This validates the fact, for chirped pulse excitation of quantum dots, the sign of the chirp determines the efficiency, due to phonon influences \cite{glassl_biexciton_2013,luker_phonon_2017}. 

Finally, we also confirm the nature of the photon emission from the quantum dot under the resonant Rabi and the chirped excitations. To this end, we perform Hanbury-Brown and Twiss ($g^{(2)}(0)$) measurements and observe that the computed $g^{(2)}(0) \approx 0.005$ (see SI), validating the single-photon nature of the emission. 

\section*{Conclusions}
To summarize, we demonstrated a systematic study of designing and fabricating compact, plug-and-play chirped fiber Bragg gratings tailored to a GaAs quantum dot resonance. Our high-efficiency femtosecond phase mask inscription method of CFBGs enabled precise control over the temporal characteristics of the excitation pulses. We established the versatility of the CFBGs by performing the two-photon chirped pulse excitation of a quantum dot under positive and negative chirps. Under positively chirped excitation, we observed a high-efficiency generation of single photons at exciton and biexciton energies, relying on the adiabatic rapid passage, while for negative chirps, we observed lower efficiency state preparation, in accordance with previous reports. In comparison with the prevailing methods for chirped pulse excitation, our method is robust, highly efficient, and optomechanically stable. Our results are in excellent agreement with theoretical simulations. In conclusion, these open an exciting avenue toward realizing an all-fiber-coupled, and alignment-free solid-state source of frequency-multiplexed single and entangled photon pairs. The versatility of CFBGs also allows for the development of efficient and compact pulse-shaping methods tailored to control various quantum states in any solid-state platform.    

\section*{Acknowledgements}
The authors gratefully acknowledge insightful discussions with Timothy Imogore (Jena), Thorsten Goebel (Jena), Dmitry Pestov (IPG Photonics), Ron Stepanek (Meadowlark Optics), and Thomas Niedereichholz (Hamamatsu). VR, RS, FK, YK and GW acknowledge financial support through the Austrian Science Fund FWF projects TAI-556N (DarkEneT), W1259 (DK-ALM Atoms, Light, and Molecules), FG5, and I4380 (AEQuDot). TKB and DER acknowledge financial support from the German Research Foundation DFG through project 428026575 (AEQuDot). AR and SFCdS acknowledge the FWF projects FG 5, P30459, I4320, the Linz Institute of Technology (LIT), the European Union's Horizon 2020 research, and innovation program under Grant Agreement Nos. 899814 (Qurope), 871130 (ASCENT+), and the QuantERA II Program (project QD-E-QKD, FFG Grant No. 891366). RGK acknowledges financial support from the German Federal Ministry of Education and Research through project 13N16028 (MHLASQU) and the German Research Foundation DFG (455425131, OH-SUPER). DR acknowledges financial support from the German Research Foundation DFG through project 448663633 (Fs$^2$CVBG).

\section*{Author contributions}
VR, GW and SN conceived and supervised the project. RGK, DR and MPS designed and fabricated the chirped fiber Bragg gratings. SFCdS and AR provided the quantum dot sample. RS, FK, YK, RGK, and VR built the setups and performed the experiments. VR and RGK wrote the first draft of the manuscript with input from other authors. FK, YK and RGK performed the data analysis and numerical simulations. All the authors discussed the results and were involved in writing the manuscript.

\clearpage

\section{Supplementary Information}
\subsection{Chirped laser pulses}
Chirped pulses possess time-varying frequency. In the frequency domain, this is described by
\begin{equation}
    \label{equ:chirp_omega}
    E(\omega)=E_{0} \exp \left[-\frac{\left(\omega-\omega_{c}\right)^{2}}{\Delta \omega^{2}}\right]\exp \left[i \frac{\phi_{2}}{2}\left(\omega-\omega_{c}\right)^{2}\right],
\end{equation}
where $E_{0}$ is the amplitude of the Gaussian frequency envelope centered at $\omega_{c}$ with a frequency bandwidth of $\Delta \omega$. The constant $\phi_{2}$ describes the group delay dispersion, GDD, or linear chirp. 

By definition, the GDD is the second derivative of the spectral phase with respect to the angular frequency $\omega$, or the first derivative of group delay ($T$) with respect to frequency. i.e., 
\begin{equation}
   \phi_2 = \text{GDD} = \dv[2]{\phi}{\omega} = \frac{\partial T}{\partial \omega} = D_{2} (\omega).
\end{equation}

In a CFBG, the time delay of wavelengths is varied by controlling the Bragg reflection resonances at various positions along the propagation direction in the fiber core. 
Converting everything to the wavelength domain, we can relate the dispersion parameter $D_{1 \lambda}$ (specified as \si{\pico\second\per\nano\meter}) to GDD (specified as \si{\pico\second\squared}) through
\begin{equation}
   D_{1\lambda} = \frac{\partial T}{\partial \lambda} = -\frac{2 \pi c}{\lambda^2} \cdot \dv[2]{\phi}{\omega} = -\frac{2 \pi c}{\lambda^2} \cdot \text{GDD} 
\end{equation}
It is now clear that a positive GDD implies a shorter time delay for the red part of the laser spectrum and vice versa for negative GDD. 

\subsection{CFBG details}
The CFBGs employed in this work are fabricated via femtosecond phase mask inscription technique \cite{thomas2007inscription,thomas2012femtosecond}
The dispersion parameter $D_{1\lambda}$ can be determined from the time delay between the reflections of various wavelength components as
\begin{equation}
    D_{1\lambda} = \frac{\Delta t} {\Delta \lambda}. 
    \label{eq_timedelay0}
\end{equation}
In the CFBG with the grating length $L$, the effective refractive index of the guided light $n_{\text{eff}}$ and $c$ as the speed of light in vacuum, the delay time is given by
\begin{equation}
    \Delta t = \frac{2 L n_{\text{eff}}} {c}. 
    \label{eq_timedelay11}
\end{equation}
Note that each wavelength component travels twice the distance within the CFBG, as it functions in reflection. Then
\begin{equation}
    D_{1\lambda} = \frac{2 L n} {c \Delta \lambda}
    \label{eq_timedelay1}
\end{equation}
which leads to
\begin{equation}
    L = \frac{c \Delta \lambda} {2 n} D_{1\lambda}
    \label{eq_timedelay2}
\end{equation}
and in terms of GDD, this will be
\begin{equation}
    L = \frac{\pi c^{2} \Delta \lambda}{n \lambda^{2}} \text{GDD},
    \label{eq_timedelay3}
\end{equation}
which provides an estimate for the required grating length to provide a given GDD. 

\begin{figure}[htb!]
	\centering
	\includegraphics[width=1\linewidth]{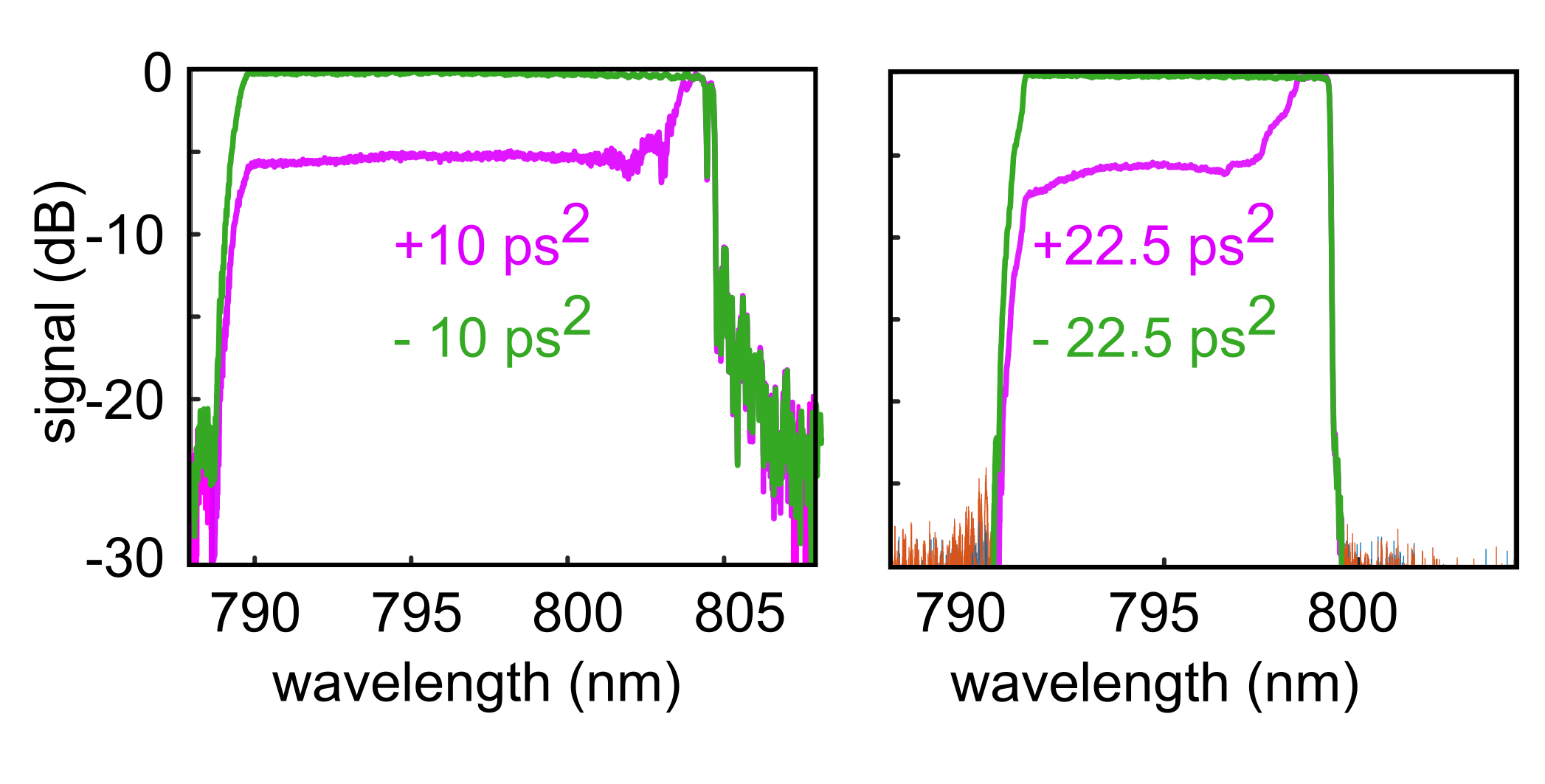}
	\caption{The measured reflectivities of \SI{10}{\pico\second\squared} and \SI{22.5}{\pico\second\squared} CFBGs.}  
 \label{fig_reflectivities}
\end{figure}

In Figure \ref{fig_reflectivities} we present the results of the further reflectivity measurements on CFBGs. Note that due to the chosen inscription pattern, for the positive GDD input side, lower wavelengths experience more losses due to the cladding mode coupling \cite{thomas2012femtosecond}. 

\begin{figure}[htb!]
	\centering
	\includegraphics[width=1\linewidth]{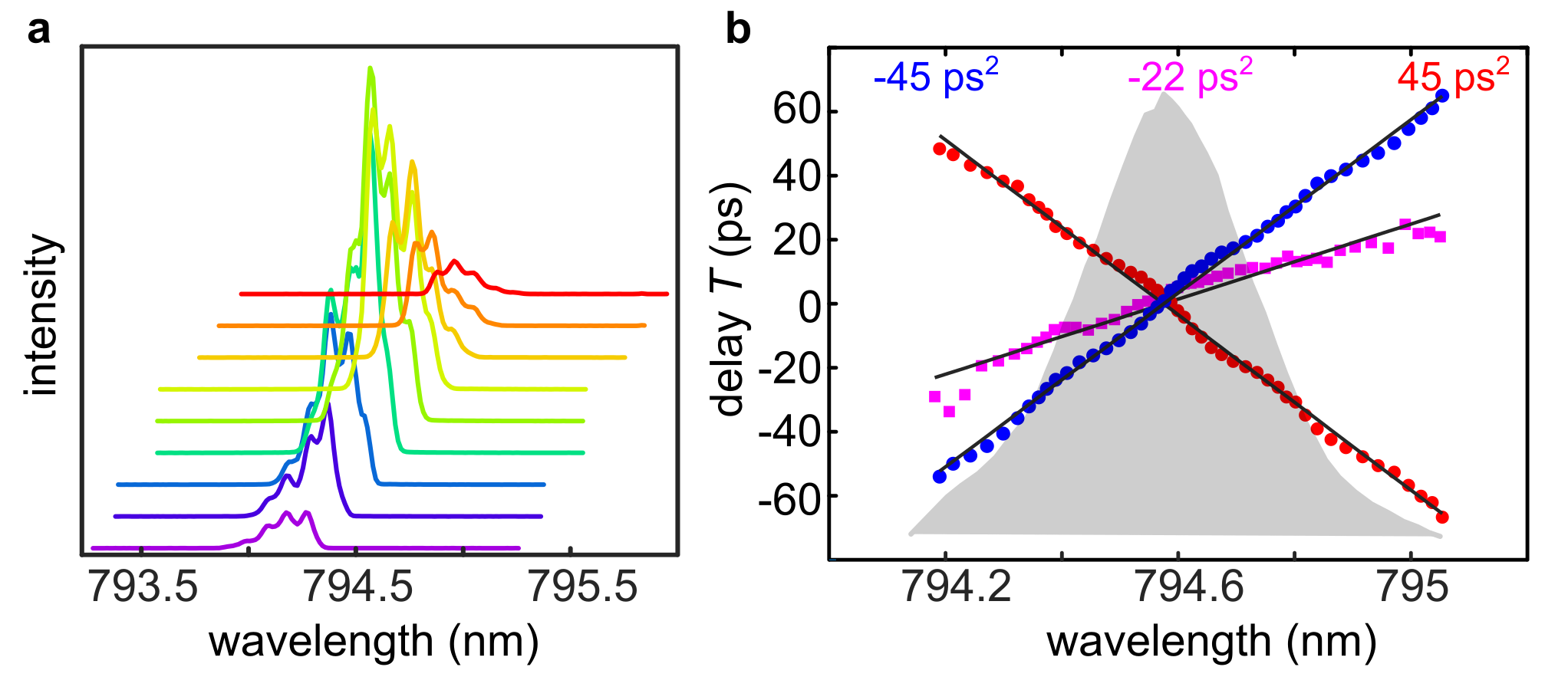}
	\caption{\textbf{Characterization of CFBG dispersions via time-of-arrival (ToA) measurement}: (a) Recorded laser spectra for various slit positions in the 4$f$ shaper (b) Results of the dispersion characterization via the ToA measurement. Grey-shaded area denotes the laser spectrum.} 
 \label{fig_tof}
\end{figure}

In the \textbf{time-of-arrival (ToA)} method of characterizing dispersion, we select several narrow spectra from the reference laser spectrum using a 4$f$ pulse shaper equipped with a motorized slit on its Fourier plane and propagate through the CFBG. The resulting, dispersed laser pulses are simultaneously recorded with a spectrometer and SNSPD (time resolution $\approx \SI{20}{\pico \second}$) using 10:90 fiber beam-splitter. While with the spectrometer we obtain the wavelength component, with the SNSPD we record its arrival time with respect to the reference laser spectrum. To extract the wavelength component precisely, we employ a Gaussian fit to the measured laser spectra, to obtain a plot of time delay as a function of central wavelength. The slope extracted from this plot provides the dispersion parameter $D_{1\lambda}$, from which the GDD can be obtained. In Figure \ref{fig_tof} we display the result of ToA measurement for $\pm\SI{45}{\pico\second\squared}$ and \SI{-22}{\pico\second\squared}.   

In Table \ref{tab:CFBGparam} we summarize the parameters of all the CFBGs and their measured performance parameters. 
\begin{table} [!hbt]
    \centering
     \setlength{\leftmargini}{0.4cm}
    \begin{tabular}{| l | c | c | c|}
        \hline
    \textbf{CFBG} & \textbf{45} & \textbf{22} & \textbf{10}\\
        \hline
    GDD$_{\text{design}}$ (\SI{}{\pico\second\squared})  & \SI{45}{} & \SI{22}{} & \SI{10}{} \\
        \hline        
    Chirp rate (\SI{}{\nano\meter\per\centi\meter}) & \SI{1.04}{} & \SI{1.54}{} & \SI{4.62}{} \\
        \hline 
    Length (\SI{}{\milli\meter}) & \SI{60}{} & \SI{60}{} & \SI{50}{} \\
        \hline 
    Center wavelength (\SI{}{\nano\meter})  & \SI{793.8}{} & \SI{795.6}{} & \SI{797.0}{} \\
        \hline  
    FWHM (\SI{}{\nano\meter})  & \SI{4.1}{} & \SI{8.1}{} & \SI{15.2}{} \\
        \hline  
    D$_{1\lambda}$ (\SI{}{\pico\second\per\nano\meter})  & \SI{-136.77}{} (\SI{135.56}{}) & \SI{+58.62}{} & -- \\
    \hline  
    GDD$_{\text{meas}}$  (\SI{}{\pico\second\squared})  & \SI{45.9}{} (\SI{-45.5}{}) & \SI{-19.7}{} & -- \\
       \hline             
    \end{tabular}
    \caption{\textbf{Overview of CFBG parameters}: The chirp rate, length, and measured full-width at half maximum (FWHM) bandwidth of the fabricated CFBGs labeled their GDD values (See SI for details).}
  \label{tab:CFBGparam}
\end{table}

\subsection{Multiple order CFBG resonances}
The change of the refractive index during the CFBG inscription process is a result of multiphoton absorption in the fiber core due to the femtosecond laser irradiation \cite{thomas2012femtosecond}. Due to the nature of this process, the underlying refractive index change is nonlinear, i.e., the modulation of the refractive index is non-sinusoidal. The pattern along the propagation axis $\Delta n(z)$ can be expressed as a sum of its Fourier terms $\Delta n_m$:
\begin{equation}
    \Delta n(z) = \frac{\Delta n_0}{2} + \sum_{m=1}^{\infty} \Delta n_m \cos\left(\frac{2\pi m}{\Lambda} z + \Phi(z)\right)
\end{equation}
with the Fourier order $m$, the nominal grating period $\Lambda$ and the grating chirp described by $\Phi(z)$. Consequently, the CFBG reflects at multiple wavelengths, corresponding to each of the Fourier terms,
\begin{equation}
    \lambda_m(z) = 2 n_{\text{eff}} \frac{\Lambda(z)}{m}.
\end{equation}
Therefore, a CFBG with a center period of \SI{545}{\nano\meter} will reflect in first order at $\lambda_1$ = \SI{1570}{\nano\meter}, in second order at $\lambda_2$ = \SI{795}{\nano\meter}, in third order at $\lambda_3$ = \SI{530}{\nano\meter} and so on.

\subsection{QD simulations}
Simulations of the pulse area-dependent biexciton population (see Figure \ref{fig_arp} in the main manuscript) are performed using a numerically complete path-integral method described in detail in Ref. \cite{cygorek2022simulation}. For this we set up a Hamiltonian consisting of the quantum dot system $\hat{H}^{\text{QD}}$, the coupling to the laser field $\hat{H}^{\text{Laser}}$ as well as the coupling to phonons $\hat{H}^{\text{Phonon}}$

\begin{equation}
    \hat{H} = \hat{H}^{\text{QD}} + \hat{H}^{\text{Laser}} + \hat{H}^{\text{Phonon}}.
\end{equation}

We model the quantum dot as a four-level system consisting of the ground state $\vert g\rangle$ two exciton states of orthogonal polarisation, $\vert x_H\rangle$ and $\vert x_V\rangle$, and the biexciton state $\vert xx\rangle$. The ground state energy is set to zero, while the two exciton states are assumed to be of the same energy $\hbar\omega_{x}$, i.e., no fine structure splitting is present. The biexciton state has a binding energy $E_B$ such that $\hbar\omega_{xx} = 2\hbar\omega_x - E_B$. 

\begin{equation}
    \begin{split}
        \hat{H}^{\text{QD}} = 
        & \hbar\omega_x \left(\vert x_H\rangle\langle x_H\vert + \vert x_V\rangle\langle x_V\vert \right)+  \hbar \omega_{xx}\vert xx\rangle\langle xx\vert 
    \end{split}
\end{equation}

Coupling to the laser field is modeled with a classical laser field 

\begin{equation}
    \begin{split}
        \hat{H}^{\text{Laser}}(t) = -\frac{\hbar}{2} \Omega(t) &( \vert g\rangle\langle x_H\vert + \vert g\rangle\langle x_V\vert \\
        &+ \vert x_H\rangle\langle xx\vert + \vert x_V\rangle\langle xx\vert + h.c. ).
        \end{split}
\end{equation}

where $\Omega(t)$ is the Fourier transform of Equation \ref{equ:chirp_omega}. In Figure \ref{fig_arp} we scale the x-axis of the simulation to be proportional to $\int_{-\infty}^{+\infty}|\Omega(t)|^2dt$, i.e., the pulse power.


Finally, we consider coupling to longitudinal-acoustic (LA) phonons via the deformation potential coupling. Here, $\hat{b}_\mathbf{k}$ ($\hat{b}^\dagger_\mathbf{k}$) annihilates (creates) a phonon of mode $\mathbf{k}$ with energy $\omega_{\mathbf{k}}$. We consider the typical pure-dephasing type coupling in the standard Hamiltonian

\begin{equation}
    \hat{H}^{\text{Phonon}} = \hbar\sum_{\mathbf{k}}\omega_{\mathbf{k}}\hat{b}^\dagger_{\mathbf{k}}\hat{b}_{\mathbf{k}} + \hbar \sum_{{\mathbf{k}},S}n_S\left( \gamma_{\mathbf{k}}^S\hat{b}_{\mathbf{k}}^\dagger + \gamma_{\mathbf{k}}^{S^*} \hat{b}_{\mathbf{k}}\right)\vert S\rangle\langle S\vert,
\end{equation}
coupling each mode $\mathbf{k}$ to the quantum dot state $\vert S\rangle$, where $S\in \{x_H,x_V,xx\}$ and $n_S$ is the number of excitations, i.e., 1 for $x_H$ and $x_H$ and 2 for $xx$.  The coupling constant $\gamma_{\mathbf{k}}^S$ and the material parameters are taken to be the same as in Ref.~\cite{Barth2016} and further simulation parameters are shown in Table \ref{tab:theoSystemParameters}.

\begin{table}
    \begin{tabular}{lc|c}
        \hline
        frequency bandwidth & $\Delta\omega$ & 1.410 $\si{\per\pico\second}$ \\
        Binding energy &$E_B$ & 4 meV\\
        QD size& $a$& 5 nm\\
        temperature & $T$ & 1.5 K
    \end{tabular}
    \caption{Overview of the parameters used for quantum dot simulation. Material parameters are taken as in Ref. ~\cite{Barth2016}.} 
        \label{tab:theoSystemParameters}
\end{table}

\subsection{Comparison of various chirping methods}
\begin{figure}[hbt!]
	\centering
	\includegraphics[width=0.9\linewidth]{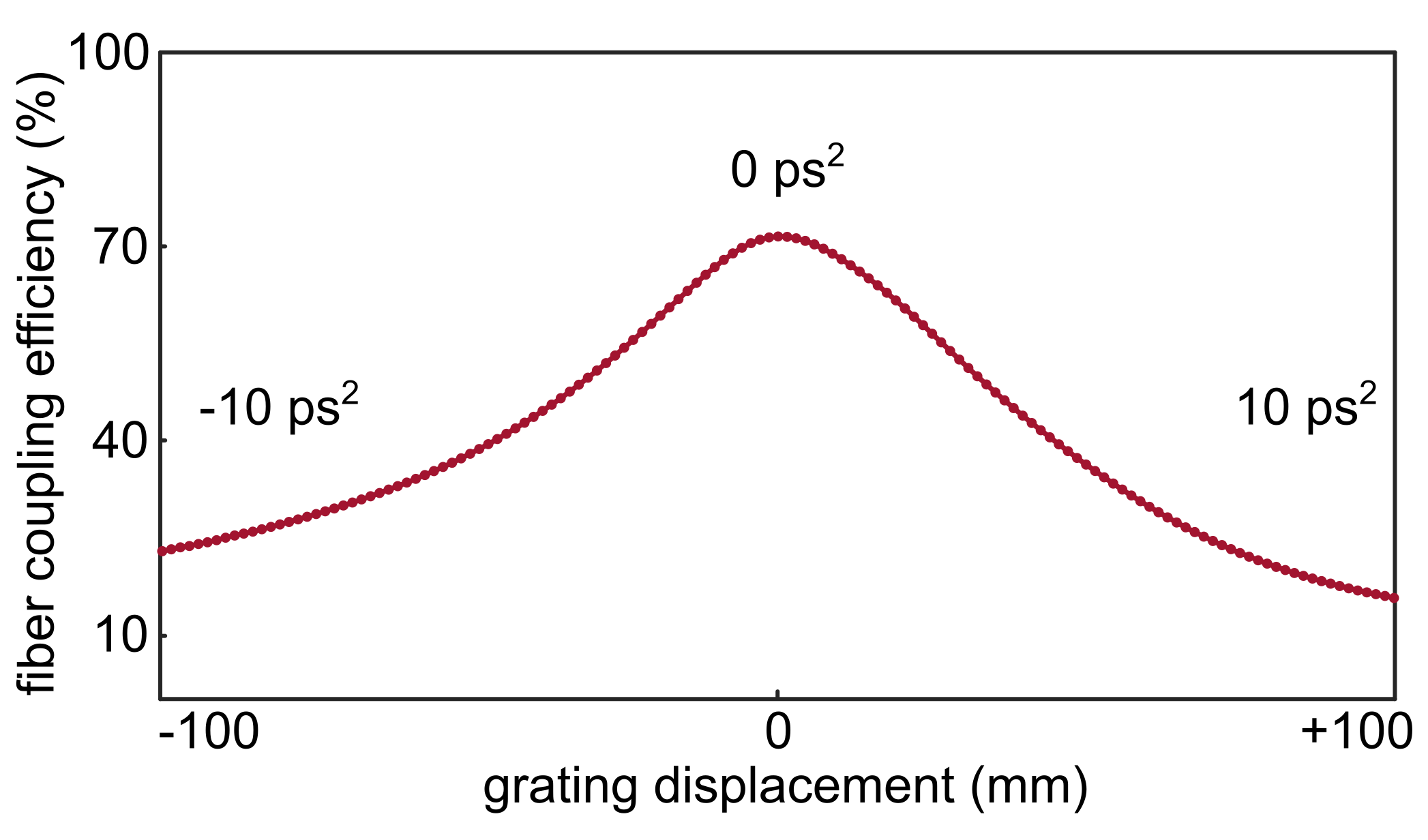}
	\caption{Variation in single-mode fiber-coupling efficiency in a grating stretcher} 
 \label{fig_fibercoupling}
\end{figure}

In this section, we compare various chirping methods. \textbf{Grating stretcher}: For imparting \SI{45}{\pico\second\squared} we designed and constructed a grating stretcher in folded geometry (details in Ref.\cite{kappe2022collective}). Here grating period $d = \SI{1200}{\per\milli\meter}$ at $\lambda_{0} = \SI{795}{\nano\meter}$ and distance $s = \SI{21}{\centi\meter}$ from the lens ($f = \SI{75}{\centi\meter}$) with angle of incidence ($\theta_{i} = \SI{2}{\degree}$) and grating efficiency is 70\%. Note that in the grating stretcher, the inherited third and fourth-order dispersions can be compensated in principle, yet, it increases the experimental complexity. The overall efficiency strongly depends on the efficiency of a single grating, i.e., $\eta\textsuperscript{4}_{\text{grating}}$. Additionally, optical aberrations also contribute to lower fiber-coupling efficiencies. To illustrate this, in Figure \ref{fig_fibercoupling} we describe the variation in single-mode fiber coupling efficiency as a function of grating position in our grating stretcher, as obtained by ray-tracing simulations.  

In an \textbf{SLM-equipped 4$f$ shapers}, the difference from the grating stretcher is that a programmable SLM is mounted at the Fourier plane. The straightforward improvement is the deterministic, adaptive, and programmable spectral phase manipulation, including higher-order components. Imparting negative dispersion does not enhance the footprint, as in grating stretchers. There are two classes of SLMs: dual-mask, transmissive SLMs \cite{mathew2014subpicosecond,remesh2019coherent,ramachandran_experimental_2021}, and two-dimensional, reflective SLMs \cite{Vaughan2005}. The \textbf{transmissive SLMs} operate by controlling the phase between frequency components, depending on the voltage applied, in a 4$f$ shaper \cite{monmayrant2010newcomer}. They suffer from transmission losses of liquid crystal layers. To perform simultaneous phase-amplitude shaping, one requires dual mask SLMs, which when operated in a folded, double-pass configuration, leads to a doubling of the transmission loss per liquid crystal layer. Therefore, the total efficiency is $ \eta\textsuperscript{4}_{\text{LC}} \cdot \eta\textsuperscript{2}_{\text{grating}}$. 
The \textbf{reflective SLM} also operates in a 4$f$ shaper, however, it functions based on diffractive pulse shaping \cite{Vaughan2005}. Here, the total efficiency is $\eta_{\text{SLM(ref)}} \cdot \eta_{\text{SLM(diff)}} \cdot \eta\textsuperscript{2}_{\text{grating}}$, where the first two terms refer to the reflection and diffraction efficiencies respectively. In a reflective SLM, typically, the total efficiency is higher than its transmissive counterpart.

In the case of \textbf{CFBG}, the maximum efficiency is $\eta\textsuperscript{2}_{\text{FOC}} \cdot \eta_{\text{CFBG}}$, where $\eta_{\text{FOC}}$ is the efficiency of the fiber-optic circulator. The characterized efficiencies are $\eta\textsuperscript{2}_{\text{FOC}} = 77\%$, and $\eta\textsuperscript{2}_{\text{CFBG}} = 90\%$. In our experiments, we observed a total efficiency of 40\%. We attribute the remaining losses to fiber mating inefficiencies. This could be mitigated by splicing the CFBG ends to the FOC.  

Generally speaking, grating stretchers (or SLM-equipped 4$f$ shapers) can be operated in a wide spectral range. Yet they have a fundamental issue: the GDD is dictated by wavelength-dependent optical path delay resulting from \textit{angular dispersion}. This imposes significantly large spatiotemporal coupling. However, the CFBG achieves spectral dispersion without angular dispersion. Thus, our CFBGs eliminate angular distortions that affect the excitation laser mode coupling to an emitter.

\subsection{QD sample}
\begin{figure*}[htb!]
	\centering
	\includegraphics[width=1\linewidth]{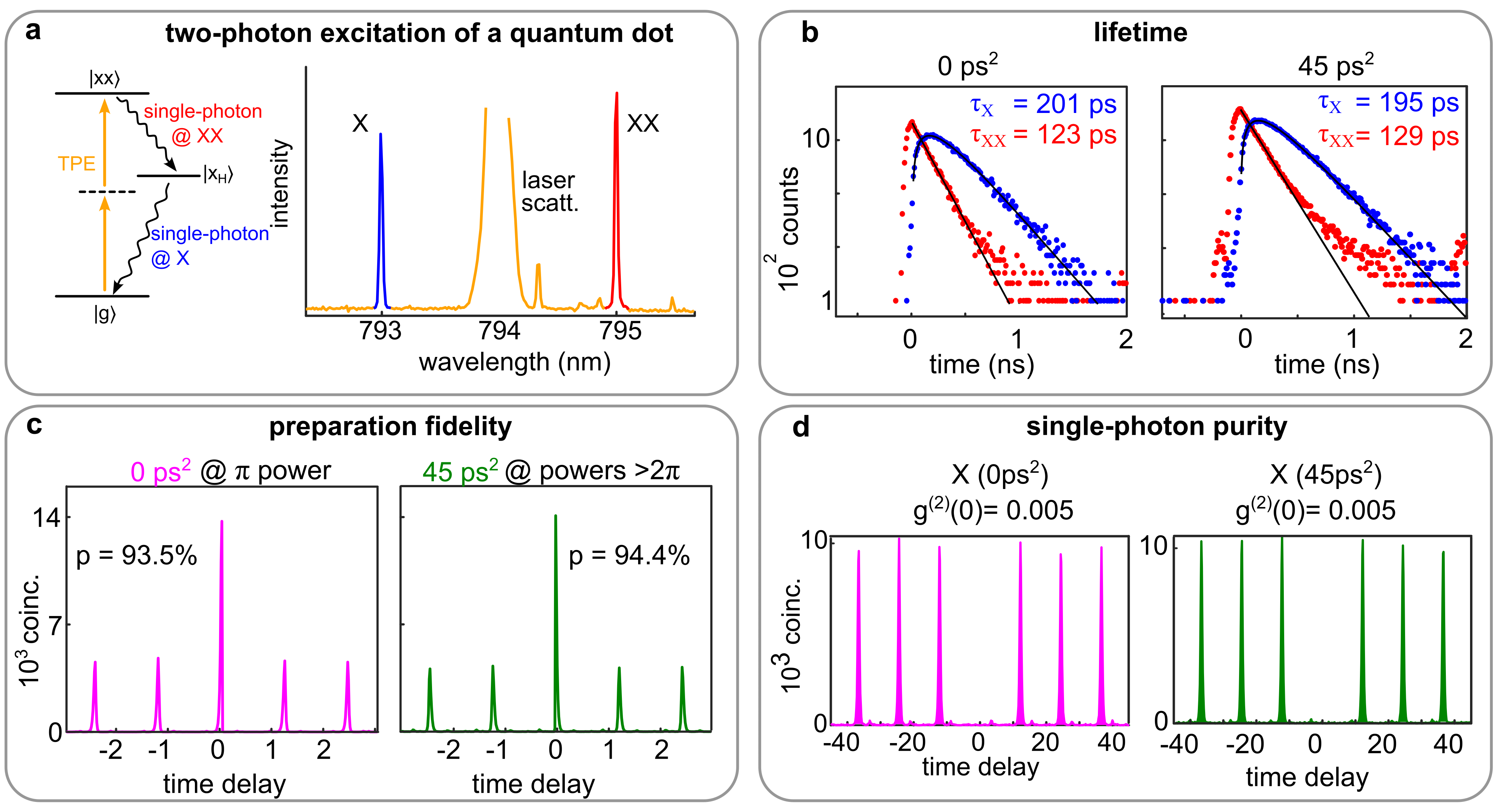}
	\caption{\textbf{Two-photon excitation and characterization}: (a) (a) Recorded emission spectrum of the quantum dot under two-photon excitation. The exciton (X) and biexciton (XX) wavelengths are identified as \SI{793.6}{\nano\meter} and \SI{795.4}{\nano\meter} respectively. (b) The recorded X and XX lifetimes under resonant Rabi and chirped excitations ($\text{GDD} = \SI{45}{\pico\second\squared}$) (c) Computed efficiencies of the biexciton state preparation under resonant Rabi and chirped excitations based on XX-X crosscorrelation measurements (d) The single-photon purity computed from the $g^{(2)}(0)$ measurements.}  
 \label{fig_photonchar}
\end{figure*}

Our sample consists of GaAs/AlGaAs quantum dots with exciton emission centered around \SI{793}{\nano\meter} grown by the Al-droplet etching method \cite{huber2017highly,da2021gaas}. The dots are embedded in the center of a lambda-cavity placed between a bottom (top) distributed Bragg reflector consisting of 9 (2) pairs of  $\lambda/4$ thick Al$_{0.95}$Ga$_{0.05}$As/Al$_{0.2}$Ga$_{0.8}$As layers. In Figure \ref{fig_photonchar} we present the characterization of the quantum dot under two-photon excitation. 

\clearpage

\bibliography{ref.bib}%

\end{document}